\documentclass[prd,twocolumn,nofootinbib,10pt,superscriptaddress]{revtex4-1}

\usepackage{amsmath,amssymb,mathtools,bbm}
\usepackage{graphicx}
\usepackage{txfonts}

\usepackage{hyperref}
\usepackage[usenames,dvipsnames]{xcolor}
\hypersetup{
colorlinks, linkcolor={blue},
citecolor={blue}, urlcolor={blue}
}

\newcommand{\Nf}{N_\mathrm{f}}
\DeclareMathOperator{\tr}{tr}
\DeclareMathOperator{\curl}{curl}

\begin{document}

\title{Confinement transition in the QED${}_3$-Gross-Neveu-XY universality class}

\author{Lukas Janssen}
\affiliation{Institut f\"ur Theoretische Physik and W\"urzburg-Dresden Cluster of Excellence ct.qmat, Technische Universit\"at Dresden, 01062 Dresden, Germany}

\author{Wei Wang}
\affiliation{Beijing National Laboratory for Condensed Matter Physics and Institute of Physics, Chinese Academy of Sciences, Beijing 100190, China}
\affiliation{School of Physical Sciences, University of Chinese Academy of Sciences, Beijing 100190, China}

\author{Michael M. Scherer}
\affiliation{Institute for Theoretical Physics, University of Cologne, 50937 Cologne, Germany}

\author{Zi Yang Meng}
\affiliation{Beijing National Laboratory for Condensed Matter Physics and Institute of Physics, Chinese Academy of Sciences, Beijing 100190, China}
\affiliation{Department of Physics and HKU-UCAS Joint Institute of Theoretical and Computational Physics, The University of Hong Kong, Pokfulam Road, Hong Kong, China}
\affiliation{Songshan Lake Materials Laboratory, Dongguan, Guangdong 523808, China}

\author{Xiao Yan Xu}
\affiliation{Department of Physics, University of California at San Diego, La Jolla, California 92093, USA}


\begin{abstract}
The coupling between fermionic matter and gauge fields plays a fundamental role in our understanding of nature, while at the same time posing a challenging problem for theoretical modeling.
In this situation, controlled information can be gained by combining different complementary approaches.
Here, we study a confinement transition in a system of $\Nf$ flavors of interacting Dirac fermions charged under a U(1) gauge field in 2+1 dimensions.
Using Quantum Monte Carlo simulations, we investigate a lattice model that exhibits a continuous transition at zero temperature between a gapless deconfined phase, described by three-dimensional quantum electrodynamics, and a gapped confined phase, in which the system develops valence-bond-solid order.
We argue that the quantum critical point is in the universality class of the QED$_3$-Gross-Neveu-XY model. 
We study this field theory within a $1/\Nf$ expansion in fixed dimension as well as a renormalization group analysis in $4-\epsilon$ space-time dimensions.
The consistency between numerical and analytical results is revealed from large to intermediate flavor number.

\end{abstract}

\maketitle

\section{Introduction}
The coupling between fermionic matter and gauge fields is of fundamental importance in both high-energy and condensed-matter physics.
In the latter context, gauge fields can emerge as a consequence of fractionalization in quantum materials.
Prominent examples include spin liquids~\cite{balents2010spin,XGWen2019} and deconfined quantum critical points~\cite{PhysRevB.70.144407,PhysRevLett.98.227202,PhysRevX.7.031052,Zhou2019Quantum,Ma2018,JGuo2019} in frustrated magnets.
As such states are characterized by topological order, it is often difficult to identify in an experiment or a simulation the relevant low-energy excitations and their characteristic properties.

In many cases, however, it is possible to tune the system by nonthermal external parameters, such as pressure or magnetic field, through a zero-temperature transition between an exotic phase with topological order and a conventionally ordered phase. If this quantum phase transition is continuous, the presence of fractionalized low-energy excitations on the topologically ordered side of the transition leaves characteristic fingerprints on the pertinent quantum critical behavior. These are in principle measurable and can be used to identify the topological order.
To take advantage of this fact, it is decisive to gain a comprehensive understanding of the quantum universality classes that involve fluctuating gauge fields.

Recent work on a system of fermions coupled to a discrete $\mathbbm{Z}_2$ gauge field in 2+1 dimensions has demonstrated the existence of such a quantum transition between a deconfined phase with gapless fermionic excitations at weak coupling and a symmetry-broken confined phase with gapped fermionic excitations at strong coupling~\cite{PhysRevX.6.041049, gazit2017emergent, gazit2018confinement, PhysRevB.96.205104,ChuangChen2019,Gonzalez2020}.
Systems of gapless fermions coupled to gauge fields with continuous gauge groups are of significant current interest as well.
An important example is (2+1)-dimensional quantum electrodynamics (QED$_3$) with U(1) gauge group, both in its compact and noncompact version 
\cite{Polyakov1975,Polyakov1977,hands2002,hands2004,armour2011,fiore2005,arakawa2005,arakawa2006,
PhysRevD.94.065026, PhysRevLett.121.041602, PhysRevD.90.036002, dipietro2016, Janssen:2016nrm, herbut2016, chester2016anomalous,
PhysRevD.94.056009, Kotikov:2016prf, gusynin2016}.
This theory is a natural candidate for the low-energy description of gapless U(1) spin liquids, which may be realizable in certain planar magnets~\cite{PhysRevLett.86.3871, PhysRevB.66.144501, PhysRevB.70.214437, PhysRevB.72.104404, PhysRevB.71.075103, PhysRevX.7.031020,XYSong2019}.

\begin{figure*}[t]
\includegraphics[width=\textwidth]{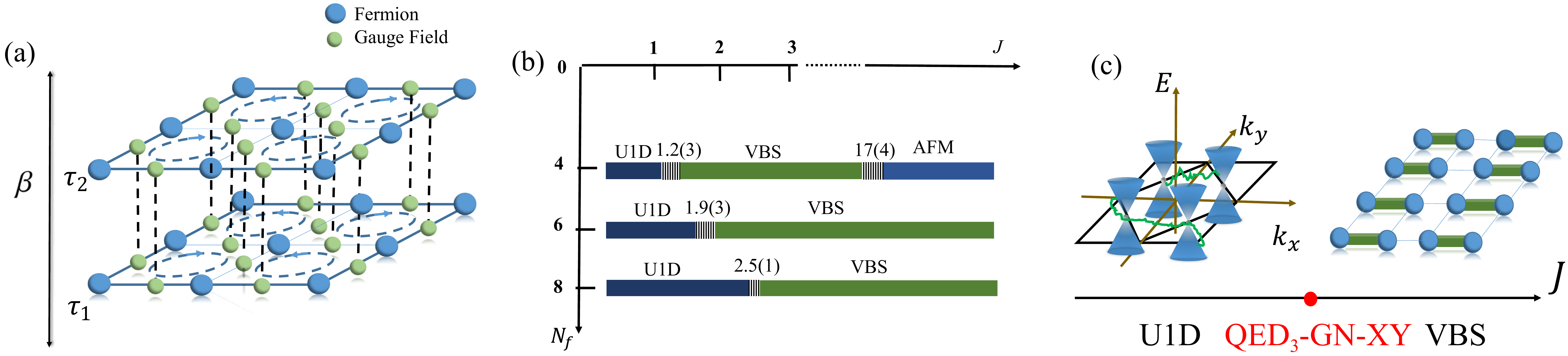}
\caption{(a) Illustration of the lattice model in Eqs.~\eqref{eq:rotor1} and \eqref{eq:rotor2}.  Blue balls represent fermions and green balls represent the U(1) gauge field, coupled to the nearest-neighbor fermion hopping. The flux term is represented by blue dashed circles in each plaquette. The gauge fields fluctuate from $\phi_{ij}(\tau_1)$ at imaginary time slice $\tau_1$ to $\phi_{ij}(\tau_2)$ at time slice $\tau_2$.
(b) Zero-temperature phase diagram as a function of the coupling $J$, parametrizing the strength of the gauge field fluctuations, for different fermion flavor numbers $\Nf$. In this work, we study the transition between the U1D phase and the confined VBS phase.
(c) Illustration of the confinement transition from U1D to VBS. In the U1D phase, the fermions form Dirac cones and interact via a U(1) gauge field, representing a lattice realization of QED$_3$. In the VBS phase, the fermions form gapped spin singlets. We argue that this transition is in the QED$_3$-Gross-Neveu-XY universality class.}
\label{fig:model}
\end{figure*}

Using Quantum Monte Carlo (QMC) simulations, the phase diagram of a lattice model of fermions coupled to a compact U(1) gauge field has recently been mapped out by some of the present authors~\cite{xu2019, PhysRevB.100.085123}. 
The model has a free parameter that can be used to drive a transition from a U(1) deconfined (U1D) phase with nodal dispersion of the fermionic excitations to confined phases in which spin and/or lattice symmetries are spontaneously broken, see Fig.~\ref{fig:model}.
The U1D phase represents a lattice realization of QED$_3$, while the confined phases describe different conventionally ordered phases.

Meanwhile, several analytical works, such as renormalization group (RG) and large-$\Nf$ calculations, have made quantitative predictions for the universal behavior of quantum critical fermion systems coupled to gauge fields~\cite{Janssen:2017eeu,Ihrig2018,PhysRevB.98.165125,
PhysRevD.98.085012,PhysRevB.99.195135,Benvenuti:2018cwd,Benvenuti:2019ujm,
Zerf2019,Dupuis:2019uhs,Dupuis:2019xdo,
Boyack:2018urb}.
However, while there has been tremendous progress in the case of the ungauged (2+1)-dimensional Gross-Neveu-like transitions~\cite{
Janssen:2014gea,PhysRevB.94.245102,PhysRevB.96.165133,PhysRevD.96.096010,PhysRevB.98.125109,Ray:2018gtp,
Gracey:2018qba,Gracey2018review,
Iliesiu:2015qra,Iliesiu:2017nrv,
Toldin:2014sxa,Otsuka:2015iba,TCLang2019,YZLiu2019,huffman2019fermion}, the situation with fluctuating gauge fields remains significantly less clear.
In particular, a quantitative comparison of universal properties between analytical predictions from the continuum quantum field theory and measurements in corresponding lattice simulations has so far not been possible.
Such a comparison is the goal of this work.

The remainder of the paper is organized as follows: 
In Sec.~\ref{sec:lattft}, we review the lattice model studied in Ref.~\cite{xu2019} and present additional evidence for a continuous transition between deconfined and confined phases.
The continuum field theory that we propose to describe the universal behavior of the confinement transition is discussed in Sec.~\ref{sec:contft}.
Section~\ref{sec:numerics} contains the comparison of the field-theory predictions with the numerical results.
Conclusions are drawn in Sec.~\ref{sec:discussion}.

\section{Lattice model and QMC results}
\label{sec:lattft}
%
We simulate a square-lattice model described by the Euclidean action $S = \int_0^\beta d\tau (L_F+L_\phi)$ with
\begin{align} 
\label{eq:rotor1}
	L_F  =  &  \sum_{\langle ij \rangle, \alpha} \psi_{i\alpha}^\dagger \left[ (\partial_\tau -\mu )\delta_{ij} - t e^{i\phi_{ij}}\right]\psi_{j\alpha} + \text{h.c.}, \allowdisplaybreaks[1] \\
	L_{\phi} = &   \frac{4}{J\Nf \Delta \tau^2}\sum_{\langle ij \rangle} \left[ 1- \cos\left(\phi_{ij}(\tau+1)-\phi_{ij}(\tau)\right)\right] \nonumber \\
	& +\frac{1}{2}K\Nf \sum_{\square}\cos(\curl \phi),
\label{eq:rotor2}
\end{align}
where $\psi^\dagger_{i\alpha}$ and $\psi_{i\alpha}$ denote fermion creation and annihilation operators at lattice site $i$ and flavor index $\alpha=1,\dots,\Nf$. The compact U(1) gauge field lives on nearest-neighbor bonds $\langle ij \rangle$ and is denoted by $\phi_{ij}$. The nearest-neighbor fermion hopping amplitude $t$ is modulated with phase $\phi_{ij}$, thereby inserting magnetic flux in each plaquette. The expression $\curl \phi$ sums the gauge fields in each elemental plaquette~$\square$ and the coupling $K>0$ stabilizes a $\pi$ flux in each plaquette, see Fig.~\ref{fig:model}(a).

In the $J \to 0$ limit, the gauge field has no imaginary time dynamics.
The ground state in this limit is characterized by gapless Dirac excitations, as both $L_F$ and the flux term in $L_\phi$ with $K>0$ favor a $\pi$ flux in each plaquette.
The Brillouin zone contains two Dirac cones per flavor and the number of (irreducible) Dirac fermions is $2\Nf$.

Turning on a small finite $J$ allows the U(1) gauge field to fluctuate. 
This model has been studied by sign-problem-free QMC simulations~\cite{xu2019}.
For $\Nf=2, 4, 6, 8$, the low-temperature phase for small $J$ has been found to be characterized by a gapless spectrum and emergent scale invariance. 
This phase represents a lattice realization of QED$_3$ with deconfined excitations and as such has been dubbed U1D.
In particular, for small $J$, the simulations suggested no evidence for confining monopole proliferation~\cite{PhysRevB.100.085123}.

Upon increasing $J$ beyond a certain threshold, however, the fermions acquire a mass, monopoles of the compact gauge field start to proliferate, and the fermions exhibit a confining potential, see Fig.~\ref{fig:model}(b). 
For $\Nf = 4,6,8$, the confined phase is described by a valence bond solid (VBS), characterized by an ordered array of spin singlets, thereby breaking lattice translation and rotation symmetries spontaneously.

\begin{figure*}[t]
\includegraphics[width=\linewidth]{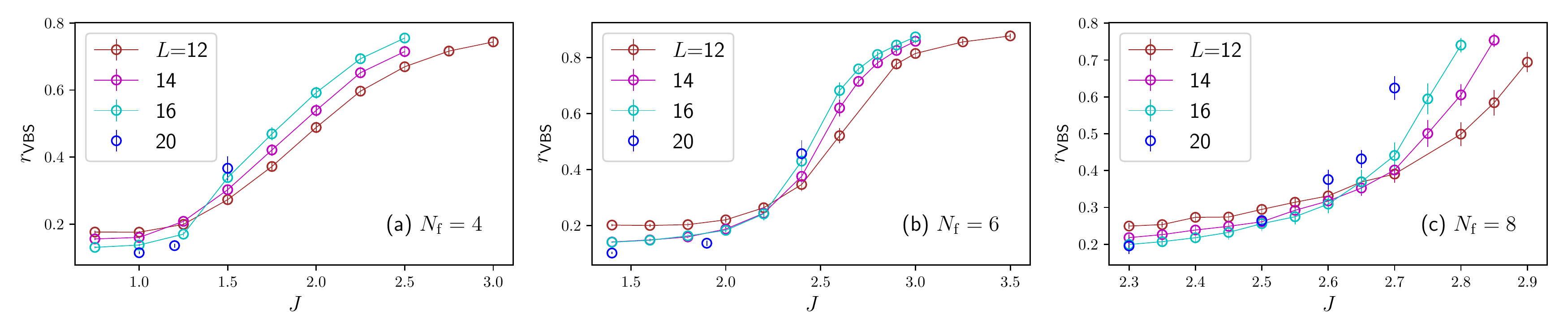}
\caption{
Correlation ratio of VBS order parameter across the U1D-VBS phase transition with linear system size up to $L=20$, for 
(a) $\Nf=4$, 
(b) $\Nf=6$, and
(c) $\Nf=8$. Correlation ratios for $\Nf=10$ ($J_c\approx 2.9$) and $\Nf=12$ ($J_c\approx 3.2$) are not shown here.
}
\label{fig:rvbs}
\end{figure*}

To improve the understanding of the U1D-VBS transition, we have carried out additional QMC simulations near the transition point and on larger lattices. We also calculated larger $\Nf$ case with $\Nf=10$ and $12$.
Technical details of the simulations, including the update scheme, the determination of the phase diagram, and the calculation of dynamical properties near the transition, are given in Refs.~\cite{xu2019,PhysRevB.100.085123}.
Fig.~\ref{fig:rvbs} shows the numerical data of VBS correlation ratios \cite{Pujari2016}
\begin{align}
\label{eq:correlationratio}
	r_{\text{VBS}} = 1- \frac{\chi_D(\vec{M}+\delta\vec{q})}{\chi_D(\vec{M})},
\end{align}
where $\vec{M}=(\pi,0)$, $\delta\vec{q}=(\frac{2\pi}{L},0)$, for different flavor numbers $\Nf$ in the vicinity of the U1D-VBS phase transition. In the above equation, $\chi_D$ denotes the VBS structure factor, given as
\begin{align}
	\chi_D(\vec{k}) = \frac{1}{L^4} \sum_{ij} \left( \langle D_iD_j \rangle -\langle D_i \rangle \langle D_j\rangle \right)e^{\text{i}\vec{k}\cdot(\vec{r}_i-\vec{r}_j)},
\end{align}
where the dimer operator $D_i=\sum_{\alpha \beta} S_\beta^\alpha(i)S_\alpha^\beta(i+\hat{x})$ is defined along the $\hat{x}$ direction, and the $S_\beta^\alpha(i)$ is the spin operator defined as $S_\beta^\alpha(i) = \psi_{i\alpha}^\dagger \psi_{i\beta} - \frac{1}{\Nf}\delta_{\alpha\beta}\sum_{\gamma}\psi_{i\gamma}^\dagger \psi_{i\gamma}$.
In the thermodynamic limit, $r_\text{VBS}$ is zero (one) in the U1D disordered (VBS ordered) phase. The curves on different lattices suggest a single crossing point in the thermodynamic limit, implying the existence of a continuous quantum phase transition.

\begin{figure}[t]
\includegraphics[width=\linewidth]{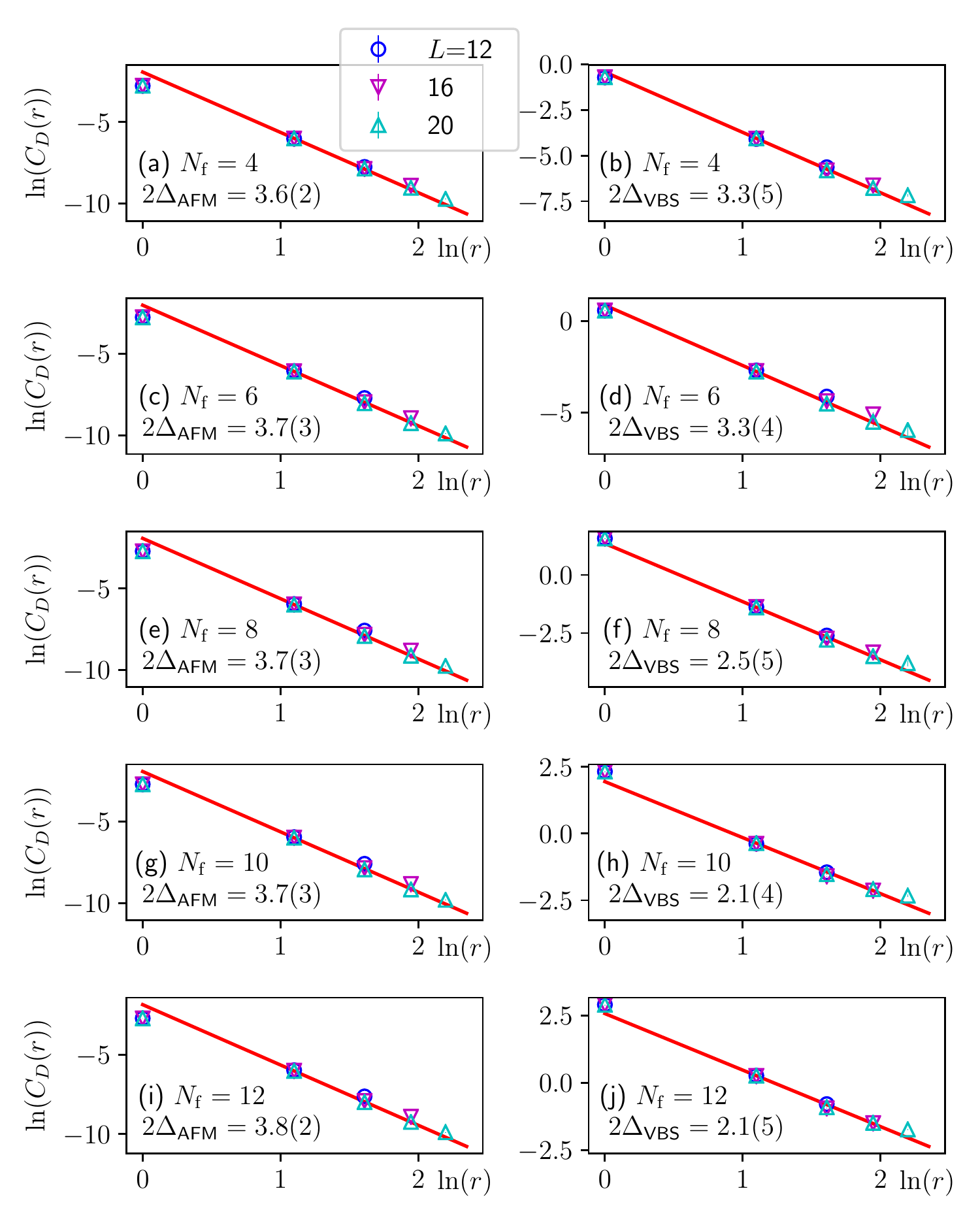}
\caption{Real space decay of spin (left column) and dimer (right column) correlation function for $\Nf=4, 6, 8, 10, 12$, at the U1D-VBS transition point. The scaling dimensions are measured by fitting the slopes in the log-log plot.
}
\label{fig:realspacedecay}
\end{figure}

More evidence for a continuous phase transition comes from the measurement of the spin and dimer correlation functions $C_S(r)$ and $C_D(r)$. Technical details concerning these measurements are deferred to Appendix~\ref{appendix}.
Fig.~\ref{fig:realspacedecay} shows the real-space decay of the spin and dimer correlators at the crossing points of the correlation ratios for different flavor numbers~$\Nf$. Instead of the usual exponential behavior, the correlators show a power-law decay with characteristic exponents $2\Delta_\text{AFM}$ and $2\Delta_{\text{VBS}}$.%
\footnote{$\Delta_{\text{AFM}}$ and $\Delta_\text{VBS}$ will be defined as scaling dimensions of AFM and VBS order parameters in Sec.~\ref{sec:contft}.}
This constitutes further strong evidence for quantum criticality beyond the findings previously presented in Ref.~\cite{xu2019}.

The VBS order parameter has a discrete $\mathbbm{Z}_4$ rotation symmetry, associated with the four possible VBS ordering patterns on the square lattice. At criticality, such a $\mathbbm{Z}_4$ symmetry is usually expected to get enhanced to a continuous $\mathrm{U}(1) \simeq \mathrm{O}(2)$ rotation symmetry~\cite{lou2007,NSMa2019}.
This suggests that the low-energy description of the quantum critical point has to include three different types of excitations: $2\Nf$ flavors of (irreducible) gapless Dirac fermions $\psi$ and $\bar\psi$, coupled to a U(1) gauge field $a_\mu$, and a real two-component scalar field $\vec \phi =  (\phi^x,\phi^y)$ serving as an order parameter for spontaneous O(2) symmetry breaking.
The minimal theory that is consistent with these requirements is the QED$_3$-Gross-Neveu-XY (QED$_3$-GN-XY) model proposed in Ref.~\cite{xu2019}.
The purpose of the rest of this paper is to confirm that the numerical data is indeed consistent with this proposal.

\section{Continuum field theory}
\label{sec:contft}
In this section, we discuss the critical behavior of the QED$_3$-GN-XY model within two complementary analytical approaches: a $1/\Nf$ expansion in fixed dimension and a RG analysis in $D=4-\epsilon$ space-time dimensions.
The Lagrangian of the QED$_3$-GN-XY model in $D=3$ reads
\begin{align}
\label{eq:QED3-GN-XY-model}
	\mathcal L & = \bar\psi_i \gamma_\mu (\partial_\mu - i a_\mu) \psi_i + \phi^a \bar\psi_i (\mu^a \otimes \mathbbm 1_2 \otimes \mathbbm 1_{\Nf/2})_{ij} \psi_j
	\nonumber \\ & \ +\!\frac{1}{2g^2} \phi^a (r-\partial_\mu^2)\phi^a\!+\!\lambda (\phi^a\phi^a)^2\!+\!\frac{1}{2 e^2} (\epsilon_{\mu\nu\rho} \partial_\nu a_\rho)^2,
\end{align}
where $\psi$ and $\bar\psi = \psi^\dagger \gamma_0$ are two-component spinors, $(\phi^a) = (\phi^x, \phi^y)$, $a=1,2$, is a real XY order-parameter field, and the $2\times 2$ Dirac matrices satisfy the Euclidean Clifford algebra $\{\gamma_\mu, \gamma_\nu\} = 2\delta_{\mu\nu} \mathbbm 1_2$, $\mu,\nu = 0,1,2$. In the Yukawa vertex $\mu^a \otimes \mathbbm 1_2 \otimes \mathbbm 1_{\Nf/2}$, the $2 \times 2$ Pauli matrices $(\mu^a) = (\mu^x, \mu^y)$ connect the two Dirac nodes, $\mathbbm 1_2$ acts on spin indices, and $\mathbbm 1_{\Nf/2}$ acts on the additional flavor indices; see Ref.~\cite{PhysRevB.72.104404} for details. Thus, the indices $i$ and $j$ run from $1$ to $2\Nf$, and we assume $\Nf$ even.
$a_\mu$ denotes the U(1) gauge field and $\epsilon_{\mu\nu\rho}$ is the totally antisymmetric tensor.
In addition to the gauge symmetry, the model features a global axial U(1) symmetry under which the two-component scalar field $(\phi^a)$ transforms as an SO(2) vector.
The parameters $e$, $g$, and $\lambda$ denote the charge, Yukawa, and quartic scalar couplings.
The parameter $r$ can be used to tune through the U1D-VBS transition.

For $r>0$, the scalar field $\phi^a$ can be integrated out and the theory describes the U1D phase with enhanced $\mathrm{SU}(2\Nf)$ symmetry.
For $r<0$, the fermions are fully gapped, since $\gamma_0\mu^a$ anticommutes with the Hamiltonian $\mathcal H(p) = \gamma_0\gamma_i p_i$, and the gauge field confines due to the proliferation of monopoles \cite{Polyakov1975,Polyakov1977,PhysRevB.70.214437,XYSong2019}. This represents the VBS ordered phase. 
The critical point is given by $r = 0$ at tree level.

\begin{figure}[t]
\includegraphics[width=\linewidth]{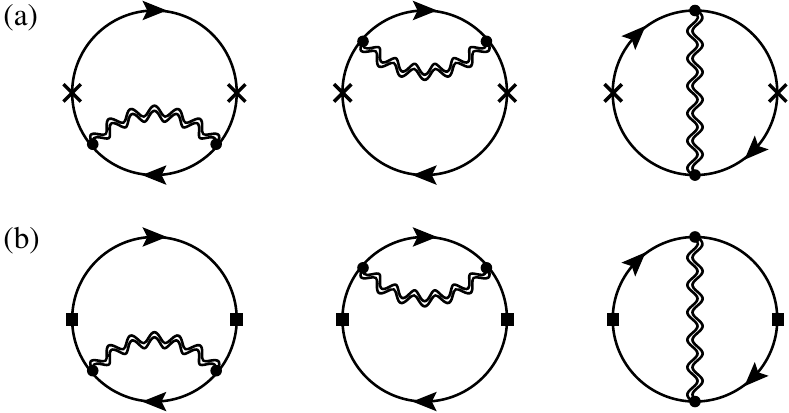}
\caption{Diagrams that contribute to the scaling dimensions $\Delta_\mathrm{AFM}$ (a) and $\Delta_\mathrm{VBS}$ (b) at $\mathcal O(1/\Nf)$ from gauge fluctuations. Solid (wiggly) lines correspond to fermion-field (gauge-field) propagators at $\mathcal O(1/\Nf^0)$. The cross $\times$ denotes $\mu^z \otimes \sigma^\alpha \otimes \mathbbm 1_{\Nf/2}$ insertions and the square $\blacksquare$ denotes $\mu^a \otimes \mathbbm 1_2 \otimes \mathbbm 1_{\Nf/2}$ insertions. Aslamazov-Larkin diagrams (not shown) vanish due to $\tr(\mu^z) = \tr(\mu^a) = 0$.}
\label{fig:diagram-gauge}
\end{figure}

Besides the universal exponents that describe the critical behavior in the vicinity of the quantum phase transition, we aim at computing the correlators of the staggered magnetization
\begin{align}
	O_\mathrm{AFM}^\alpha(\vec r) \sim (-1)^{r_x + r_y} S^\alpha({\vec r}) \sim \bar\psi_i(\mu^z \otimes \sigma^\alpha \otimes \mathbbm 1_{\Nf/2})_{ij} \psi_j,
\end{align}
where $\alpha = x,y,z$ denotes the spin components and $\vec r = r_x \hat x + r_y \hat y$ the lattice site,
and the staggered dimer operator
\begin{multline}
	O^a_\mathrm{VBS}(\vec r) \sim \left((-1)^{r_y} \vec S({\vec r}) \cdot \vec S({\vec r + \hat y}), (-1)^{r_x} \vec S({\vec r}) \cdot \vec S({\vec r + \hat x})\right)^a
	\nonumber \\
	\sim \bar\psi_i (\mu^a \otimes \mathbbm 1_2 \otimes \mathbbm 1_{\Nf/2} )_{ij} \psi_j \sim \phi^a
\end{multline}
with $a = x,y$ \cite{PhysRevB.72.104404}.
In the above equations, two operators $\mathcal O$ and $\mathcal O'$ are considered equivalent, $\mathcal O \sim \mathcal O'$, if they transform in the same way under all symmetries.
At criticality, the correlation functions are given by power laws
\begin{align}
	\langle O^\alpha_\mathrm{AFM}(\vec r) \, O^\alpha_\mathrm{AFM}(0) \rangle & \propto \frac{1}{r^{2\Delta_\mathrm{AFM}}},\\
	\langle O^a_\mathrm{VBS}(\vec r) \, O^a_\mathrm{VBS}(0) \rangle & \propto \frac{1}{r^{2\Delta_\mathrm{VBS}}},
\end{align}
where $\Delta_\mathrm{AFM}$ and $\Delta_\mathrm{VBS}$ are the scaling dimensions of the operators $O_\mathrm{AFM}^\alpha$ and $O_\mathrm{VBS}^a$.
The correlation function of a microscopic lattice operator will in general be a linear combination of all operators in the conformal field theory that are equivalent to the respective lattice operator. In the long-wavelength limit, however, the operator with the lowest scaling dimension will dominate.

\subsection{Large-$\Nf$ expansion}
\label{sec:contft-A}
%

\begin{figure}[t]
\includegraphics[width=\linewidth]{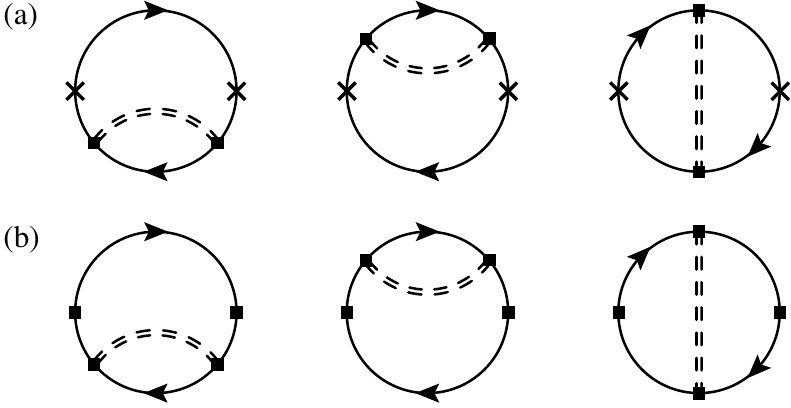}
\caption{Diagrams that contribute to the scaling dimensions $\Delta_\mathrm{AFM}$~(a) and $\Delta_\mathrm{VBS}$~(b) at $\mathcal O(1/\Nf)$ from scalar fluctuations. Dashed lines correspond to scalar-field propagators at $\mathcal O(1/\Nf^0)$.}
\label{fig:diagram-scalar}
\end{figure}

In the limit of large fermion flavor number $\Nf$, the universal critical exponents and the scaling dimensions of the fermion bilinears can be computed analytically.
We start by discussing the scaling dimensions $\Delta_\mathrm{AFM}$ and $\Delta_\mathrm{VBS}$.
The diagrams that contribute at order $\mathcal O(1/\Nf)$ are shown in Figs.\ \ref{fig:diagram-gauge} and \ref{fig:diagram-scalar}. We note that at this order of the $1/\Nf$ expansion, mixed diagrams that involve both scalar and gauge field propagators are finite and therefore do not contribute to the scaling dimensions. This can be understood as a consequence of the Ward identity associated with the U(1) gauge symmetry, in analogy to the QED$_3$-Gross-Neveu-Ising case~\cite{PhysRevD.98.085012,PhysRevB.99.195135}.

\begin{figure}[t]
\includegraphics[width=\linewidth]{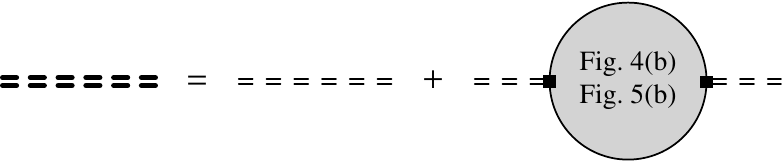}
\caption{Dyson equation for the scalar-field two-point correlator at~$\mathcal O(1/\Nf)$. Thick dashed line corresponds to scalar-field propagator at $\mathcal O(1/\Nf)$.}
\label{fig:dyson}
\end{figure}

The gauge-field contributions to the scaling dimensions (Fig.~\ref{fig:diagram-gauge}) are well known \cite{PhysRevB.72.104404,PhysRevB.66.144501,PhysRevB.99.195135,chester2016anomalous},
\begin{align}
	\Delta_\mathrm{AFM}\Bigr|_{\mathrm{QED_3}} = \Delta_\mathrm{VBS}\Bigr|_{\mathrm{QED_3}} = 2 - \frac{32}{3\pi^2 \Nf} + \mathcal O(1/\Nf^2).
\end{align}
We note that Aslamazov-Larkin diagrams vanish due to $\tr(\mu^z) = \tr(\mu^a) = 0$~\cite{PhysRevB.99.195135}.
Both operators have exactly the same scaling dimensions as a consequence of the enhanced $\mathrm{SU}(2\Nf)$ symmetry for $r>0$ in the U1D phase~\cite{xu2019, PhysRevB.66.144501}.
However, at criticality, $r=0$, this symmetry is explicitly broken and the scalar-field fluctuations lead to different values for $\Delta_\mathrm{VBS}$ and $\Delta_\mathrm{AFM}$. 
The scalar contributions to the AFM bilinear are shown in Fig.~\ref{fig:diagram-scalar}, yielding
\begin{align} 
 \label{eq:delta-AFM-Scenario2}
	\Delta_\mathrm{AFM}\Bigr|_{\text{GN-XY}} & = 2 - \frac{8}{3\pi^2\Nf} + \mathcal O(1/\Nf^2).
\end{align}
The lowest-dimensional operator in the conformal field theory with the symmetries of the staggered dimer operator is the scalar field $\phi^a$.
Its scaling dimension can be computed using the Dyson equation, see Fig.~\ref{fig:dyson} and Ref.~\cite{PhysRevB.99.195135}.
The scalar contributions are shown in Fig.~\ref{fig:diagram-scalar}(b), yielding
\begin{align}
\label{eq:delta-VBS-Scenario2}
	\Delta_\mathrm{VBS}\Bigr|_{\text{GN-XY}} = \Delta_\phi \Bigr|_{\text{GN-XY}} = 1 - \frac{4}{3\pi^2\Nf} + \mathcal O(1/\Nf^2).
\end{align}
Since $\Delta_\phi = (D-2+\eta_\phi)/2$, we obtain the order-parameter anomalous dimension of the pure Gross-Neveu-XY model (i.e., without the coupling to a gauge field) as
\begin{align}
	\eta_\phi\Bigr|_\text{GN-XY} = 1 - \frac{8}{3\pi^2\Nf} + \mathcal O(1/\Nf^2), \label{eq:eta-phi-GN-XY}
\end{align}
which is consistent with the previous calculation~\cite{hands1993}.

Including both gauge-field and scalar-field contributions yields the scaling dimensions in the QED$_3$-GN-XY model as
\begin{align}
\label{eq:delta-AFM-Scenario3}
	\Delta_\mathrm{AFM}\Bigr|_{\text{QED$_3$-GN-XY}} & = 2 - \frac{40}{3\pi^2\Nf} + \mathcal O(1/\Nf^2), \\
\label{eq:delta-VBS-Scenario3}
	\Delta_\mathrm{VBS}\Bigr|_{\text{QED$_3$-GN-XY}} & = 1 + \frac{28}{3\pi^2\Nf} + \mathcal O(1/\Nf^2),
\end{align}
with the order-parameter anomalous dimension as
\begin{align}
	\eta_\phi\Bigr|_\text{QED$_3$-GN-XY} = 1 + \frac{56}{3\pi^2\Nf} + \mathcal O(1/\Nf^2). \label{eq:eta-phi-QED3-GN-XY}
\end{align}
We note that the two-point correlator of the fermion bilinear $\bar\psi_i(\mu^a \otimes \mathbbm{1}_2 \otimes \mathbbm 1_{\Nf/2})_{ij} \psi_j$ vanishes to all orders in the large-$\Nf$ calculation, which can be understood as a consequence of the fact that this operator becomes zero when integrating out $\phi$ in the critical large-$\Nf$ theory \cite{Benvenuti:2018cwd, Benvenuti:2019ujm, PhysRevB.99.195135}.%
\footnote{We are grateful to J.\ Maciejko for pointing this out to us.}
The correlation-length exponent $\nu$ can be obtained within an analogous calculation to that of Ref.~\cite{PhysRevB.99.195135} for the QED$_3$-Gross-Neveu-Ising case. We find
\begin{align} \label{eq:nu-QED3-GN-XY}
	\nu^{-1}\Bigr|_\text{QED$_3$-GN-XY} = 1 - \frac{80}{3\pi^2\Nf} + \mathcal O(1/\Nf^2),
\end{align}
in agreement with the recent calculation presented in Ref.~\cite{Boyack:2019joi}.

In order to make contact with the $4-\epsilon$ expansion result discussed in the next subsection, it is instructive to generalize Eqs.~\eqref{eq:eta-phi-QED3-GN-XY}, \eqref{eq:nu-QED3-GN-XY} to general space-time dimension $2<D<4$. Using the integration formulae derived in Refs.~\cite{hands1993, PhysRevD.98.085012}, we obtain
\begin{multline}
	\eta_\phi\Bigr|_\text{QED$_3$-GN-XY} = 4 - D
	\\
	+ \frac{4(D^2-2)\Gamma(D-1)}{D^2(D-2)\Gamma(-D/2)\Gamma(D/2)^3} \frac{1}{\Nf} + \mathcal O(1/\Nf^2)
\end{multline}
and
\begin{multline}
	\nu^{-1} \Bigr|_\text{QED$_3$-GN-XY} = D - 2
	\\
	+ \frac{2(D^2-2D+2)\Gamma(D)}{D(D-2)\Gamma(1-D/2)\Gamma(D/2)^3} \frac{1}{\Nf} + \mathcal O(1/\Nf^2),
\end{multline}
in agreement with Ref.~\cite{Gracey:1993ka}.
Further expanding the above equations in $\epsilon = 4 - D$ leads to
\begin{align} \label{eq:large-N-etaphi}
	\eta_\phi\Bigr|_\text{QED$_3$-GN-XY} & = \epsilon + \frac{7\epsilon}{2\Nf} - \frac{9\epsilon^2}{8\Nf} + \mathcal O(\epsilon^3,1/\Nf^2), \\
	\label{eq:nunivlargeNf}
	\nu^{-1} \Bigr|_\text{QED$_3$-GN-XY} & = 2 - \epsilon - \frac{15 \epsilon}{2 \Nf} + \frac{41 \epsilon^2}{8 \Nf} + \mathcal O(\epsilon^3,1/\Nf^2),
\end{align}
which, as we show below, agrees with the leading-order result that we obtain within the $4-\epsilon$ expansion. This constitutes another nontrivial cross-check of our calculations.

The above analysis neglects the compactness of the U(1) gauge field in our lattice model.
A compact gauge field admits instanton events, which correspond to insertion of magnetic flux in a localized region of space-time. In fact, the monopole operators, which perform this flux insertion, are relevant when the fermions are gapped out, leading to confinement of charges in the VBS phase \cite{Polyakov1975,Polyakov1977,PhysRevB.70.214437,XYSong2019}.
In the gapless U1D phase, the scaling dimension $\Delta_{q=1/2}$ of the least irrelevant monopole operator with minimal magnetic charge $q=1/2$ (corresponding to $2\pi$ flux insertion) can be computed by employing the operator-state correspondence, which implies that $\Delta_{q=1/2}$ is given by the ground-state energy of the corresponding theory on a sphere in the presence of a magnetic monopole \cite{Borokhov:2002ib}.
For pure QED$_3$, this computation gives
\begin{align}
	 \Delta_{q=1/2}\Bigr|_\text{QED$_3$} = c_\text{QED$_3$} \cdot 2\Nf + \mathcal O(1/\Nf^0).
\end{align}
with $c_\text{QED$_3$} = 0.265\ldots$ 
(The correction at order $\mathcal O(1/\Nf^0)$ has been computed in~\cite{Pufu:2013vpa}.)
We note that $\Delta_{q=1/2} \propto \Nf$ in the large-$\Nf$ limit. The technical reason for this property is that the fermions can be integrated out in this limit, such that $\Delta_{q=1/2}$ is simply the ground-state energy of $\Nf$ free Dirac fermions in the presence of a classical background gauge field.
At large enough $\Nf$, monopoles are therefore irrelevant and the ground state of QED$_3$ hence describes a stable U1D phase with conformal invariance.%
\footnote{
For small $\Nf$, the scaling dimension $\Delta_{q=1/2}$ may drop below $D=3$, such that monopoles would become relevant. The critical value of $\Nf$, below which this happens, cannot be computed in a controlled way within the $1/\Nf$ expansion.}
The same argument applies to the QED$_3$-GN-XY model. Hence,
\begin{align}
	 \Delta_{q=1/2}\Bigr|_\text{QED$_3$-GN-XY} = c_\text{QED$_3$-GN-XY} \cdot 2\Nf + \mathcal O(1/\Nf^0).
\end{align}
The constant $c_\text{QED$_3$-GN-XY}$ can be computed along the lines of Refs.~\cite{Dupuis:2019uhs, Dupuis:2019xdo}. In fact, this calculation turns out to be agnostic towards the number of scalar boson components,%
\footnote{We thank W.\ Witczak-Krempa for pointing this out to us.}
such that the leading-order monopole scaling dimension of the  QED$_3$-GN-XY model agrees with that of the QED$_3$-Gross-Neveu-Ising model, $c_\text{QED$_3$-GN-XY} = c_\text{QED$_3$-GN-Ising} = 0.195\dots$ \cite{Dupuis:2019uhs}.
In particular, $\Delta_{q=1/2} > D = 3$ for large enough $\Nf$, such that all monopoles operators are irrelevant and the critical theory indeed describes a stable RG fixed point with a unique infrared relevant direction (corresponding to the tuning parameter of the continuous transition). 

\begin{figure}[t]
\includegraphics[width=\linewidth]{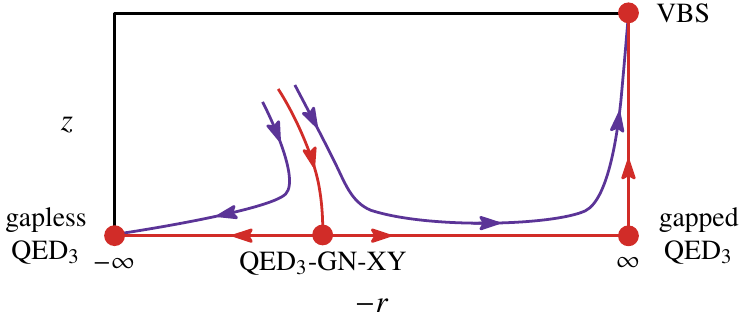}
\caption{Schematic RG flow in the space spanned by tuning parameter $r$ and monopole fugacity $z$ for large $\Nf$. The theory defined by Eq.~\eqref{eq:QED3-GN-XY-model} describes the line $z=0$. Both in the U1D phase, described by the infrared stable fixed point denoted as gapless QED$_3$, and at the QED$_3$-GN-XY quantum critical point, monopole operators are (dangerously) irrelevant. However, when the fermions acquire a mass gap, the fugacity $z$ grows towards the infrared, indicating a proliferation of monopoles and confinement. The corresponding fixed point, denoted as gapped QED$_3$, is unstable towards the formation of VBS order.}
\label{fig:flow}
\end{figure}

The emerging picture resembles the situation at the deconfined N\'eel-VBS critical point in quantum magnets, which is described by the noncompact CP$^1$ model~\cite{PhysRevB.70.144407,senthil2004}, see Fig.~\ref{fig:flow}.
In the gapless U1D phase (N\'eel phase in case of quantum magnets), which is described by QED$_3$ (symmetry-broken phase of the noncompact CP$^1$ model), monopoles are irrelevant in the large-$\Nf$ limit, and we can neglect the compactness of the U(1) gauge field.
The same applies to the quantum critical point, described by the QED$_3$-GN-XY model (noncompact CP$^1$ model) at criticality.
However, when the fermions acquire a mass gap in the ordered phase of the QED$_3$-GN-XY model (corresponding to the paramagnetic phase of the noncompact CP$^1$ model), monopoles become relevant, indicating confinement and VBS order \cite{Polyakov1975,Polyakov1977,read1989,PhysRevB.70.214437,XYSong2019}.
As a consequence, on the VBS side of the transition, two independent diverging length scales are present, just as in the deconfined N\'eel-VBS critical point in quantum magnets~\cite{PhysRevB.70.144407}. In addition to the correlation length $\xi$ associated with the flow of the tuning parameter, a longer length scale $\xi'$ that is associated with the proliferation of monopoles exists.
Observation of $\xi'$ in a QMC simulation is in principle possible, but requires significantly larger lattices~\cite{shao2016}. We leave this for future work.

\subsection{RG analysis in $D=4-\epsilon$ dimensions}
\label{sec:contft-B}
%
We now demonstrate the existence of the quantum critical QED$_3$-GN-XY fixed point beyond the $1/\Nf$ expansion within a standard RG analysis.
The QED$_3$-GN-XY theory can be straightforwardly generalized to general space-time dimension $2<D<4$ by replacing the $2\Nf$ flavors two-component Dirac fermions by $\Nf$ flavors of four-component fermions~\cite{Gehring:2015vja}. 
In agreement with the QED$_3$-Gross-Neveu-Ising case~\cite{Janssen:2017eeu, Ihrig2018, PhysRevB.98.165125}, all couplings become simultaneously marginal in four space-time dimensions. This allows to set up an $\epsilon$ expansion around four space-time dimensions in the same way as in the standard Wilson-Fisher $\phi^4$ theory.  
Integrating over the $D$-dimensional momentum shell $\Lambda/b< \sqrt{\omega^2+\vec{q}^2} <\Lambda$ with cutoff $\Lambda$ and $b>1$ causes the couplings to flow according to the equations
\begin{align}
	\frac{d e^2}{d \ln b} & = \epsilon e^2-\frac{4}{3} \Nf e^4, \\
	\frac{d g^2}{d \ln b} & = \epsilon g^2 - (\Nf+1)g^4 + 6 e^2 g^2, \\
	\frac{d \lambda}{d \ln b} & = \epsilon\lambda - 2 \Nf g^2(\lambda-g^2)-5\lambda^2,
\end{align}
where we have rescaled $(e^2, g^2, \lambda)\mapsto (e^2, g^2, \lambda)/(4\pi^2 \Lambda^{\epsilon})$ and assumed that the theory is tuned to the critical point.
Note that these flow equations agree with those of the continuum description of the Kekul\'e transition on the honeycomb lattice in the presence of a U(1) gauge field~\cite{PhysRevB.94.205136}.
They are also equivalent to those of Ref.~\cite{Dupuis:2019uhs} upon setting their $N_b$ to 2.
The flow admits an infrared stable RG fixed point at
\begin{align}
	(e^{2}_\ast, g^2_\ast, \lambda_\ast) & = \left(\frac{3}{4 \Nf}, \frac{2\Nf+9}{2\Nf(\Nf+1)}, \frac{C(\Nf) - 8 - \Nf}{10(\Nf+1)} \right) \epsilon 
	\nonumber \\ & 
	\quad + \mathcal O(\epsilon^2),
\end{align}
with $C(\Nf) = \sqrt{\Nf^2+56\Nf+424+810/\Nf}$.

The wave function renormalization function of the two-component scalar field reads
\begin{align}
	\gamma_\phi(g^2)& = \Nf g^2.
\end{align}
This expression is formally identical to the corresponding equation in the non-gauged Gross-Neveu-XY model, because the XY scalar field is not charged.
At the fixed point, we obtain the order parameter anomalous dimension as
\begin{align}
	\eta_\phi&= \gamma_\phi(g^2_\ast) = \frac{2\Nf+9}{2(\Nf+1)} \epsilon + \mathcal O(\epsilon^2).
\end{align}
Further expanding this result for large $\Nf$ leads to
\begin{align}
	\eta_\phi&=\epsilon + \frac{7\epsilon}{2 \Nf} 
	- \frac{7\epsilon}{2 \Nf^2} 
	+ \mathcal O(\epsilon^2,1/\Nf^3),
\end{align}
in agreement with Eq.~\eqref{eq:large-N-etaphi}.
The correlation-length exponent $\nu$ can be obtained from the flow of the tuning parameter,
\begin{align}
	\frac{d r}{d \ln b}& = (2-2\lambda-\Nf g^2) r + 2\lambda - 2\Nf g^2,
\end{align}
implying
\begin{align}
	\nu^{-1} = 2 - \frac{2 C(\Nf) + 8 \Nf + 29}{10 (\Nf + 1)}\epsilon + \mathcal O(\epsilon^2).
\end{align}
To next-to-leading order in $1/\Nf$, we obtain
\begin{align}
	\nu^{-1} = 2 - \epsilon - \frac{15 \epsilon}{2\Nf} 
	+\frac{87 \epsilon}{2 \Nf^2}
	+ \mathcal O(\epsilon^2,1/\Nf^3),
\end{align}
in agreement with Eq.~\eqref{eq:nunivlargeNf}, as announced above.

\begin{figure*}[t]
\includegraphics[width=\textwidth]{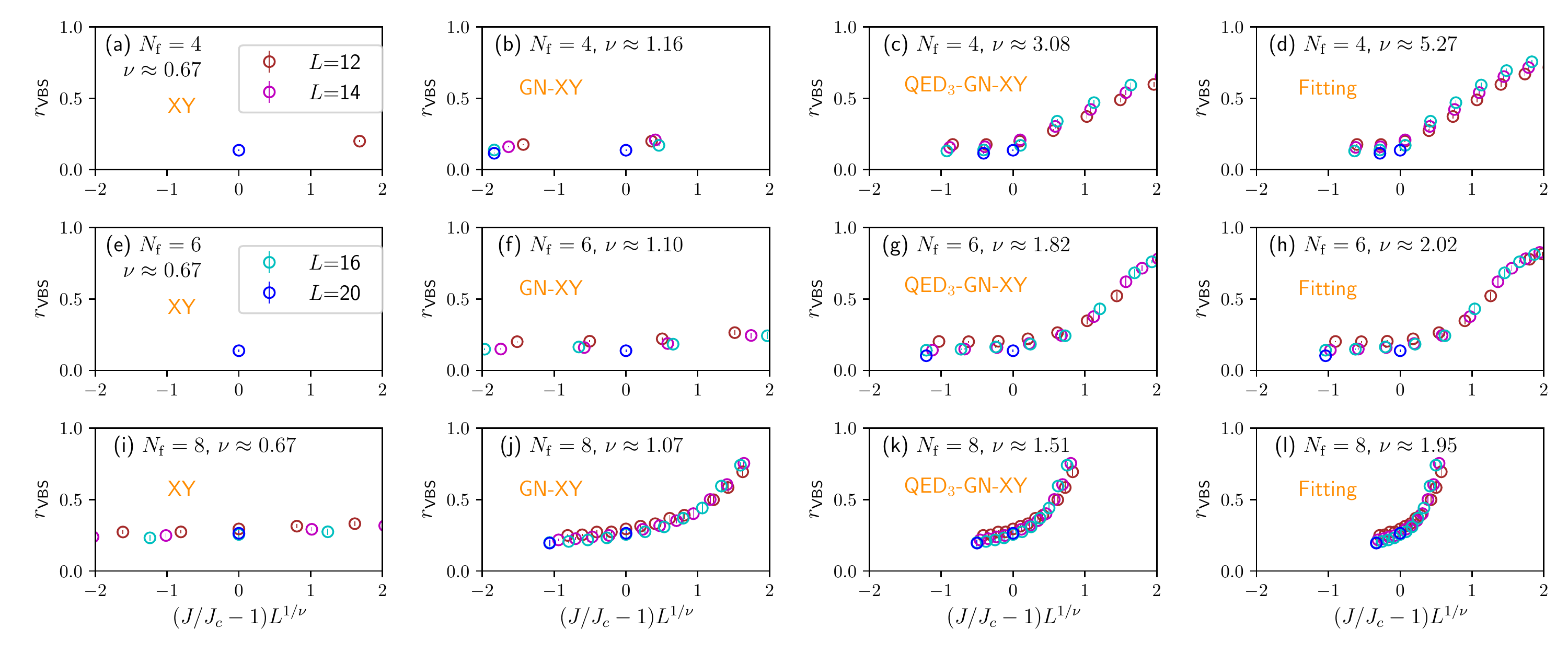}
\caption{Data collapses of VBS correlation ratios according to Eq.~\eqref{eq:eq37} within the three different scenarios (first three columns) and using fitted values (last column) for $\Nf=4$ (top row), $\Nf = 6$ (middle row), and $\Nf=8$ (bottom row).
(a,e,i)~Scenario 1, assuming standard XY universality.
(b,f,j)~Scenario 2, assuming pure Gross-Neveu-XY universality.
(c,g,k)~Scenario 3, assuming QED$_3$-GN-XY universality.
(d,h,l)~Using fitted values as denoted in the insets.}
\label{fig:nf468collapse}
\end{figure*}

\section{Evidence for QED$_3$-GN-XY universality}
\label{sec:numerics}
%
We now discuss three possible universality classes that could in principle describe the continuous U1D-VBS transition observed in the lattice model.
Among these three scenarios, we will see that the numerical data is consistent only with the QED$_3$-GN-XY universality class.

The transition between the U1D disordered phase and the VBS ordered phase is detected via a real two-component order parameter.
Hence, if the U1D-VBS transition were a standard quantum transition within the Landau paradigm, quantum-to-classical mapping would apply, and one would expect the transition to be in the classical 3D XY universality class. Within Scenario~1 (Scen1), we thus assume the critical exponents~\cite{Poland:2018epd,Chester2019}
%
%
\begin{align} \label{eq:scenario1}
	\text{Scen1:} &&
	\nu^{-1}\Bigr|_\text{XY} & = 1.48864(22), &
%
	\eta_\phi\Bigr|_\text{XY} & = 0.0385(7).
\end{align}

However, it is well known that the presence of gapless fermions at the transition can change the universality class~\cite{sachdevbook}. 
Previous studies in systems with Dirac fermions (but without a fluctuating gauge field) coupled to a $\mathrm O(2)$ order parameter have put forward the Gross-Neveu-XY universality class~\cite{TCLang2013,ZCZhou2016,Li:2017sbk,Xu:2018kekule,Otsuka:2018kcb, Liao:2019vbs,  Li:2019hhu, PhysRevD.96.096010}. This describes our Scenario 2 (Scen2), with exponents
%
%
\begin{align} \label{eq:scenario2}
	\text{Scen2:} &&
	\nu^{-1}\Bigr|_\text{GN-XY} & = 1 - \frac{16}{3\pi^2\Nf} + \mathcal O(1/\Nf^2), \allowdisplaybreaks[1] \\&&
	\eta_\phi\Bigr|_\text{GN-XY} & = 1 - \frac{8}{3\pi^2\Nf} + \mathcal O(1/\Nf^2), \allowdisplaybreaks[1] \\&&
	\Delta_\mathrm{VBS}\Bigr|_{\text{GN-XY}} & = 1 - \frac{4}{3\pi^2\Nf} + \mathcal O(1/\Nf^2), \allowdisplaybreaks[1] \\&&
	\Delta_\mathrm{AFM}\Bigr|_{\text{GN-XY}} & = 2 - \frac{8}{3\pi^2\Nf} + \mathcal O(1/\Nf^2),
\end{align}
cf.\ Sec.~\ref{sec:contft}.
We note in particular that $\Delta_\mathrm{VBS}\Bigr|_{\text{GN-XY}} < 1$.
Below, we will show that this result is inconsistent with our numerical data, which we ascribe to the presence of gapless gauge-field excitations at the transition.
We will demonstrate that the data is instead consistent, within error bars, with our theoretical predictions for the QED$_3$-GN-XY universality class. The universal exponents within this scenario, dubbed Scenario 3 (Scen3), have been computed in Sec.~\ref{sec:contft},
\begin{align}  \label{eq:scenario3}
	\text{Scen3:} &&
	\nu^{-1}\Bigr|_\text{QED$_3$-GN-XY} & = 1 - \frac{80}{3\pi^2\Nf} + \mathcal O(1/\Nf^2), \allowdisplaybreaks[1] \\ &&
	\eta_\phi\Bigr|_\text{QED$_3$-GN-XY} & = 1 + \frac{56}{3\pi^2\Nf} + \mathcal O(1/\Nf^2), \allowdisplaybreaks[1] \\&&
	\Delta_\mathrm{VBS}\Bigr|_{\text{QED$_3$-GN-XY}} & = 1 + \frac{28}{3\pi^2\Nf} + \mathcal O(1/\Nf^2), \allowdisplaybreaks[1] \\&&
	\Delta_\mathrm{AFM}\Bigr|_{\text{QED$_3$-GN-XY}} & = 2 - \frac{40}{3\pi^2\Nf} + \mathcal O(1/\Nf^2).
\end{align}
Note that now $\Delta_\mathrm{VBS}\Bigr|_{\text{QED$_3$-GN-XY}} > 1$.

In the vicinity of the quantum critical point at coupling $J=J_\mathrm{c}$, the dimensionless correlation ratio $r_\text{VBS}$ should obey the finite-size scaling law
\begin{align}
\label{eq:eq37}
	r_\text{VBS}(J,L) = \mathcal F((J/J_\mathrm{c} - 1) L^{1/\nu})
\end{align}
with scaling function $\mathcal F(x)$ and linear system size $L$.
Fig.~\ref{fig:nf468collapse} shows the correlation ratios $r_\text{VBS}$ as function of the rescaled coordinate $x \coloneqq (J/J_\mathrm{c} - 1) L^{1/\nu}$ for $\Nf=4,6,8$, using the predictions for $1/\nu$ within the three scenarios proposed above, as well as using fitted values.
Due to the limited system sizes that are available within tenable simulation times, the data collapses are not perfect in all cases. However, for all flavor numbers checked, the data collapses using the standard XY prediction are significantly worse than for the other two scenarios.
This rules out Scenario 1, revealing that the transition evades the usual quantum-to-classical mapping.
The differences between the qualities of the data collapses in Scenario 2 and 3 are less significant, in particular for $\Nf=8$. The reason for this behavior is that the critical exponents in these two scenarios approach the same limiting value for large $\Nf$, $\lim_{\Nf \to \infty} (1/\nu) = 1$.
Nevertheless, for $\Nf=4$ and $6$, we note that the spread of values of $r_\text{VBS}$ within the ranges of $x$ shown is significantly larger within Scenario 3 than for the other two scenarios.
We interpret this as first evidence for QED$_3$-GN-XY universality (Scenario 3).

\begin{figure}[t]
\includegraphics[width=\linewidth]{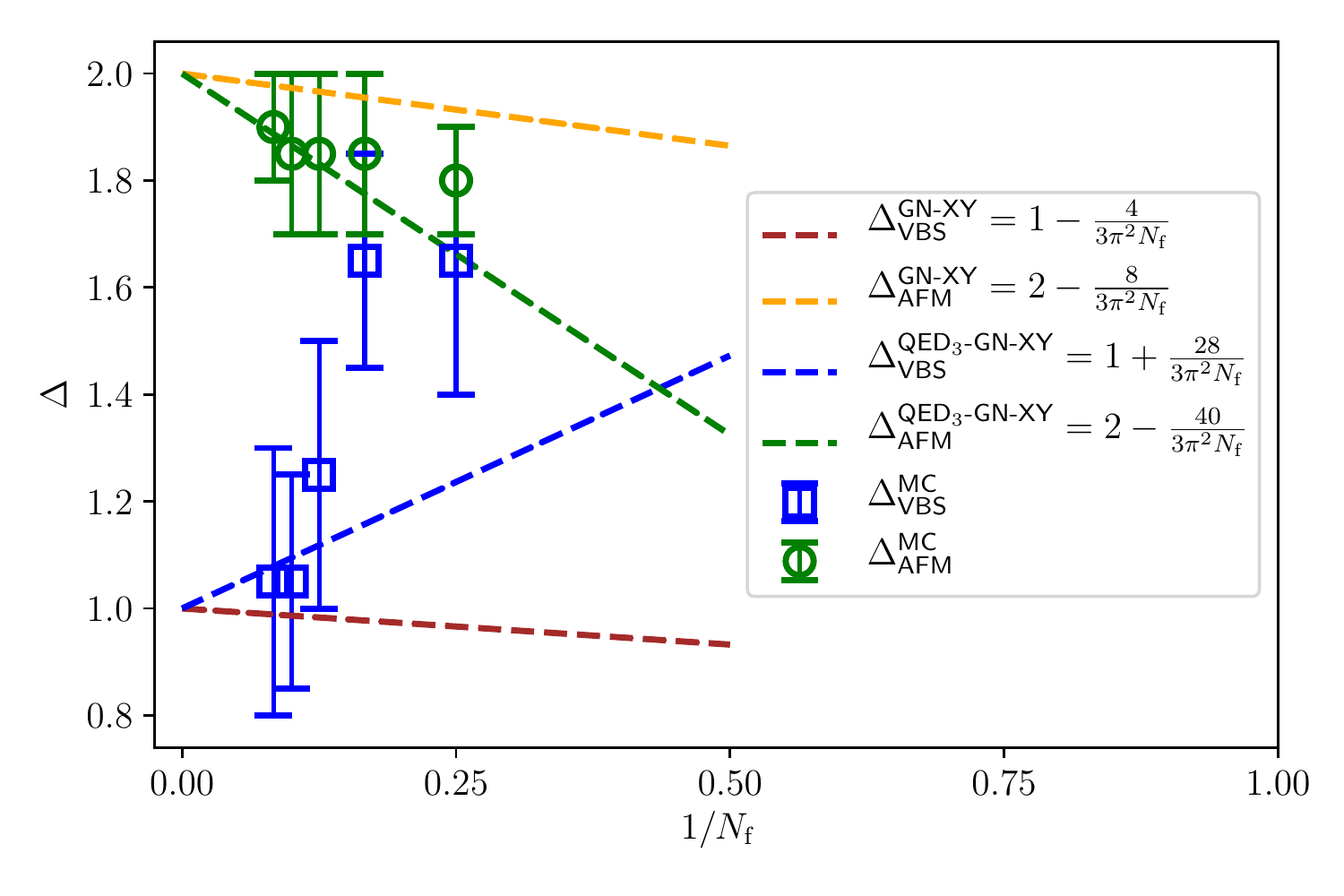}
\caption{Scaling dimensions $\Delta_\text{AFM}$ and $\Delta_\text{VBS}$ as extracted from the QMC data of Fig.~\ref{fig:realspacedecay}, in comparison the predictions of the Gross-Neveu-XY and QED$_3$-GN-XY models, respectively.  The fact that $\Delta_\text{VBS}$ is above $1$ is consistent only with the QED$_3$-GN-XY universality class (Scenario 3).
}
\label{fig:scalingdimension}
\end{figure}

To clearly rule out the pure Gross-Neveu-XY universality class (Scenario 2), we have performed fits to the dimer $C_D(r)$ and spin $C_S(r)$ correlation functions, assuming power-law behaviors
\begin{align}
	C_{D}(r) & \propto \frac{1}{r^{2\Delta_\text{VBS}}}, &
	C_{S}(r) \propto \frac{1}{r^{2\Delta_\text{AFM}}},
\end{align}
with exponents $\Delta_\text{VBS}$ and $\Delta_\text{AFM}$. 
The fits are shown as straight lines in the log-log plots of Fig.~\ref{fig:realspacedecay}, and we plot the extracted exponents in Fig.~\ref{fig:scalingdimension} in comparison with the theoretical predictions for Gross-Neveu-XY universality (Scenario 2) and QED$_3$-GN-XY universality (Scenario 3).
While the errors are large, the overall behavior is significantly closer to the QED$_3$-GN-XY predictions, in particular for larger flavor numbers $\Nf \geq 8$. We also note that we find $\Delta_\text{VBS} > 1$ for all considered values of $\Nf$. This rules out Scenario 2, suggesting that the transition observed in the simulation is indeed described by a new universality class in which the fluctuating U(1) gauge field plays a significant role.

\section{Conclusions}
\label{sec:discussion}
%
In conclusion, we have provided evidence for the existence of an unconventional quantum critical point between a deconfined phase, characterized by gapless fermionic degrees of freedom coupled to a U(1) gauge field, and a conventionally ordered phase with spontaneously broken XY symmetry, in which the fermions acquire a band gap and the gauge field confines.
The critical point can hence be understood as an example of a continuous confinement transition in $2+1$ space-time dimensions.

Using a finite-size scaling analysis of our QMC data, we have demonstrated that this transition evades the usual quantum-to-classical mapping. We have also shown that the numerical data is inconsistent with the pure Gross-Neveu-XY universality class.
Instead, the data is consistent with our predictions for a novel universality class, dubbed QED$_3$-GN-XY, which includes besides the fluctuating fermionic degrees of freedom also the effects of the coupling to the U(1) gauge field.

We have shown that the corresponding QED$_3$-GN-XY field theory can be studied within a large-$\Nf$ expansion in fixed dimension, or alternatively within an $\epsilon$ expansion in $D=4-\epsilon$ space-time dimensions.
Using these expansion schemes, we have computed estimates for the characteristic exponents that describe the QED$_3$-GN-XY universality class, including the order-parameter anomalous dimension $\eta_\phi$ and the correlation-length exponent $\nu$, as well as the scaling dimensions of the spin and dimer correlation functions.
The accuracy of these estimates increases for larger flavor number, and should be reasonable for our simulations of the cases $\Nf=6$ and $8$. (E.g., in the pure Gross-Neveu-XY universality class, the accuracy of the leading-order $1/\Nf$ results is expected to be of the order of $\lesssim 3\%$ for $\Nf= 8$~\cite{Gracey:2018qba, PhysRevD.96.096010}.)
For $\Nf=4$, the corrections may be more significant. 
This could be one of the reasons for the reduced quality of the data collapse in this case.
Improving on this would require to extend our analytical calculations to the next order in the large-$\Nf$ expansion (e.g., along the lines of the calculation in Ref.~\cite{PhysRevD.98.085012} for the QED$_3$-Gross-Neveu-Ising case) and/or to a higher order in the $4-\epsilon$ expansion (e.g., along the lines of Refs.~\cite{Ihrig2018, PhysRevB.98.165125}).
On the numerical side, refined quantitative predictions for the critical exponents and universal scaling dimensions require simulations on significantly larger lattices.
Besides an improved characterization of the universality class, such simulations could also facilitate a search of the predicted emergent second length scale $\xi'$, associated with the proliferation of monopoles on the VBS side of the transition.

\paragraph*{Note added.} During the review process of this work, a related preprint appeared~\cite{Zerf:2020mib}, which extends the calculation of the critical exponents of the QED$_3$-GN-XY universality class to $\mathcal O(1/\Nf^2)$ in the large-$\Nf$ expansion and to $\mathcal O(\epsilon^4)$ in the $4-\epsilon$ expansion. Their leading-order results agree with ours.

\begin{acknowledgments}
We thank B.\ Ihrig, J.\ Maciejko, S.\ Ray, W.\ Witczak-Krempa, and A.\ Vishwanath for valuable discussions.
LJ acknowledges support by the Deutsche For\-schungs\-ge\-mein\-schaft (DFG) through the Emmy Noether program (\mbox{JA2306/4-1}, project id 411750675), SFB 1143 (project id 247310070), and the W\"urzburg-Dresden Cluster of Excellence \textit{ct.qmat} (EXC 2147, project id 390858490).
WW and ZYM acknowledge support from the Ministry of Science and Technology of China through the National Key Research and
Development Program (Grant No.\ 2016YFA0300502), the National Science Foundation of China (Grant Nos.\ 11574359, 11674370), and Research Grants Council of Hong Kong Special Administrative Region of China (Grant No.\ 17303019).
MMS was supported by the DFG through SFB 1238 (projects C02 and C03, project id 277146847).
We thank the Center for Quantum Simulation Sciences in the Institute of Physics, Chinese Academy of Sciences, the Computational Initiative at the Faculty of Science at the University of Hong Kong and the Tianhe-1A platform at the National Supercomputer Center in Tianjin and the Tianhe-2 platform at the National Supercomputer Center in Guangzhou for technical support and generous allocation of CPU time.
\end{acknowledgments}

\vspace{\baselineskip}

\appendix

\section{Measurement of correlation functions}
\label{appendix}
We measure connected correlation functions in real space. The spin correlator is defined as
\begin{equation}
C_S(r) =  \sum_{\alpha \beta} \left\langle S_\beta^\alpha(r) S_\alpha^\beta(0) \right\rangle - \sum_{\alpha \beta} \left\langle S_\beta^\alpha(r)\right\rangle \left\langle S_\beta^\alpha(0)\right\rangle,
\end{equation}
and the dimer correlator is
\begin{equation} \label{eq:dimer-correlator}
C_D(r) = \left\langle D(r)D(0) \right\rangle - \langle D(r)\rangle \langle D(0) \rangle.
\end{equation}
For the spin correlator, the background is zero, $\langle S_\beta^\alpha(r)\rangle = 0$, which is a consequence of the unbroken SU(2) spin symmetry.
However, for the dimer correlator, the background may be finite, $\langle D(r)\rangle \ne 0$, and must be measured numerically.
Near the quantum critical point and for large distances $r$, the difference between the two terms in the right-hand-side of Eq.~\ref{eq:dimer-correlator} may become significantly smaller than their individual magnitudes, such that the dimer correlator has a larger uncertainty.


\end{document}